\documentclass[english,12pt]{article}
\usepackage[cp1251]{inputenc}
\usepackage{babel}
\usepackage{amsfonts, amsmath}
\newcommand{\be}{\begin{equation}} \newcommand{\ee}{\end{equation}}

\begin{document}
\title{\bf Deformed Density Matrix and Quantum Entropy of the Black Hole
} \thispagestyle{empty}

\author{A.E.Shalyt-Margolin\hspace{1.5mm}\thanks
{Fax +375 172 926075; e-mail:a.shalyt@mail.ru;alexm@hep.by}}
\date{}
\maketitle
 \vspace{-25pt}
{\footnotesize\noindent  National Center of Particles and High
Energy Physics, Bogdanovich Str. 153, Minsk 220040, Belarus\\
{\ttfamily{\footnotesize
}}

\rm\normalsize \vspace{0.5cm}
\begin{abstract}
In the present work the approach - density matrix deformation -
earlier developed by the author to study a quantum theory of the
Early Universe (Planck's scales) is applied to study a quantum
theory of black holes. On this basis the author investigates the
information paradox problem, entropy of the black hole remainders
after evaporation, and consistency with the holographic principle.
The possibility for application of the proposed approach to the
calculation of quantum entropy of a black hole is considered.
\end{abstract}

\section{Introduction.
Deformed  Density Matrix in QMFL}
Quantum entropy of a black hole is commonly considered as a
formula for the communal entropy representing a series, where the
major term is coincident with Bekenstein-Hawking entropy in a
semiclassical approximation, whereas other terms are its quantum
corrections. This paper presents the development of a new approach
to a quantum theory close to the singularity (Early Universe),
whose side product is the application to a quantum theory of black
holes the calculations of entropy including. The principal method
of this paper is deformation of a quantum-mechanical density
matrix in the Early Universe. And a quantum mechanics of the Early
Universe is considered as Quantum Mechanics with Fundamental
Length (QMFL), the associated deformed density matrix being
referred to as a density pro-matrix. The deformation is understood
as an extension of a particular theory by inclusion of one or
several additional parameters in such a way that the initial
theory appears in the limiting transition. In Section 1 the
formalism of the density pro-matrix in QMFL is described in detail.
In Section 2 the entropy density matrix on the
unit minimum area is introduced for different observers. Then it is used in a
detailed study of the information problem of the Universe, and in
particular, for the information paradox problem. This problem is
reduced to comparison of the initial and final densities of
entropy for one and the same observer. It's shown that according
to the natural standpoint, there is no information loss at the
closed Universe. Based on the proposed approach, in Section 3 the
quantity of the entropy for Planck's remainders of black holes is
analyzed in case when the latter are incompletely evaporated, in
an effort to give an answer for the recent J.Bekenstein's
question. Besides, consideration is given to the coordination
between the obtained results and the holographic principle. The
last Section is devoted to the possibility for application of the
proposed approach to calculation of the black hole entropy.
\\Besides, in Section 1 the principal features of QMFL construction
using the deformed density matrix are briefly outlined
 \cite{shalyt4}--\cite{shalyt20}. It is suggested that in quantum
gravitation, similar to a quantum theory of the Early Universe,
the introduction of the fundamental length $l_{min}$ is a must, as
follows from the generalized uncertainty relations
\cite{r2}--\cite{Magg} and not only \cite{r1}. Then, as noted
earlier(e.g., see \cite{shalyt5}), the fundamental length may be
included into quantum mechanics by the use of the density matrix
deformation. Recall the main features of the associated
construction. We begin with the Generalized Uncertainty Relations
(GUR) \cite{r2}:
\begin{equation}\label{U2N}
\triangle x\geq\frac{\hbar}{\triangle p}+\alpha^{\prime}
L_{p}^2\frac{\triangle p}{\hbar}.
\end{equation}
\noindent Here $L_{p}$ is the Planck's length:
$L_{p}=\sqrt\frac{G\hbar}{c^3}\simeq1,6\;10^{-35}\;m$ and
 $\alpha^{\prime} > 0$ is a constant. In \cite{r3} it was shown
that this constant may be chosen equal to 1. However, here we will
use $\alpha^{\prime}$ as an arbitrary constant without giving it
any definite value. Equation (\ref{U2N})  is identified as the
Generalized Uncertainty Relations in Quantum Mechanics.
\\The inequality (\ref{U2N}) is quadratic in $\triangle p$:
\begin{equation}\label{U3N}
\alpha^{\prime} L_{p}^2({\triangle p})^2-\hbar \triangle x
\triangle p+ \hbar^2 \leq0,
\end{equation}
from whence the fundamental length is
\begin{equation}\label{U4N}
\triangle x_{min}=2\sqrt\alpha^{\prime} L_{p}.
\end{equation}
Since in what follows we proceed only from the existence of
fundamental length, it should be noted that this fact was
established apart from GUR as well. For instance, from an ideal
experiment associated with Gravitational Field and Quantum
Mechanics a lower bound on minimal length was obtained in
\cite{Ng1}, \cite{Ng2} and  improved in \cite{Baez} without using
GUR to an estimate of the form $\sim L_{p}$. As reviewed
previously in \cite{r1}, the fundamental length appears quite
naturally at Planck scale, being related to the
quantum-gravitational effects.\noindent Let us  consider equation
(\ref{U4N}) in some detail.  Squaring both its sides, we obtain
\begin{equation}\label{U5N}
(\overline{\Delta\widehat{X}^{2}})\geq 4\alpha^{\prime} L_{p}^{2},
\end{equation}
Or in terms of density matrix
\begin{equation}\label{U6N}
Sp[(\rho \widehat{X}^2)-Sp^2(\rho \widehat{X})]\geq
4\alpha^{\prime} L_{p}^{2 }=l^{2}_{min}>0,
\end{equation}
where $\widehat{X}$ is the coordinate operator. Expression
(\ref{U6N}) gives the measuring rule used in well-known quantum
mechanics QM. As distinct from QM, however, in the are considered
here the right-hand side of (\ref{U6N}) can not be brought
arbitrary close to zero as it is limited by $l^{2}_{min}>0$, where
because of GUR $l_{min} \sim L_{p}$.
\\As demonstrated in \cite{shalyt4},\cite{shalyt5},\cite{shalyt13s},
\cite{shalyt20},QMFL may be considered as deformation of QM, the
density matrix being the principal object of this deformation and the
deformation parameter being dependent on the measuring
scale. This means that in QMFL $\rho=\rho(x)$, where $x$ is the
scale, and for $x\rightarrow\infty$  $\rho(x) \rightarrow
\widehat{\rho}$, where $\widehat{\rho}$ is the density matrix in
QM. Since from \cite{shalyt5},\cite{shalyt13s}, \cite{shalyt20}
it follows that at Planck's scale $Sp[\rho]<1$, then for such
scales $\rho=\rho(x)$, where $x$ is the scale, is not a density
matrix as it is generally defined in QM. On Planck's scale
$\rho(x)$ is referred to as "density pro-matrix". As follows from
the above, the density matrix $\widehat{\rho}$ appears in the
limit \cite{shalyt4},\cite{shalyt5}:
\begin{equation}\label{U12N}
\lim\limits_{x\rightarrow\infty}\rho(x)\rightarrow\widehat{\rho},
\end{equation}
when  QMFL turns to QM. Thus, on Planck's scale the density matrix
is inadequate to obtain all information about the mean values of
operators. A "deformed" density matrix (or pro-matrix) $\rho(x)$
with $Sp[\rho]<1$ has to be introduced because a missing part of
information $1-Sp[\rho]$ is encoded in the quantity
$l^{2}_{min}/a^{2}$, whose specific weight decreases as the scale
$a$ expressed in  units of $l_{min}$ is going up.
 In the notation
system used for $\alpha = l_{min}^{2 }/x^{2 }$, where $x$ is the
scale for the fundamental deformation parameter.

\noindent {\bf Definition 1.} (Quantum Mechanics with Fundamental
Length)

\noindent Any system in QMFL is described by a density pro-matrix
of the form $\rho(\alpha)=\sum_{i}\omega_{i}(\alpha)|i><i|$, where
\begin{enumerate}
\item $0<\alpha\leq1/4$.
\item The vectors $|i>$ form a full orthonormal system.
\item $\omega_{i}(\alpha)\geq 0$, and for all $i$  the
finite limit $\lim\limits_{\alpha\rightarrow
0}\omega_{i}(\alpha)=\omega_{i}$ exists.
\item
$Sp[\rho(\alpha)]=\sum_{i}\omega_{i}(\alpha)<1$,
$\sum_{i}\omega_{i}=1$.
\item For every operator $B$ and any $\alpha$ there is a
mean operator $B$ depending on  $\alpha$:\\
$$<B>_{\alpha}=\sum_{i}\omega_{i}(\alpha)<i|B|i>.$$
\end{enumerate}
Finally, in order that our definition 1 be in agreement with the
result of (\cite{shalyt4}, Section 2), the following condition
must be fulfilled:
\begin{equation}\label{U1}
Sp[\rho(\alpha)]-Sp^{2}[\rho(\alpha)]\approx\alpha.
\end{equation}
Hence we can find the value for $Sp[\rho(\alpha)]$ satisfying the
condition of definition 1:
\begin{equation}\label{U2}
Sp[\rho(\alpha)]\approx\frac{1}{2}+\sqrt{\frac{1}{4}-\alpha}.
\end{equation}
As regards {\bf Definition 1.}, some explanatory remarks are
needed. Of course, any theory may be associated with a number of
deformations. In case under consideration the deformation is
"minimal" as only the probabilities are deformed rather than the
state vectors.  This is essential for the external form of the
density pro-matrix, and also for points 2 and 3 in {\bf Definition
1.}. This suggests point 5 of the Definition: deformation of the
average values of the operators. And point 4 follows directly from
point 3,(\ref{U12N}) and remark before this limiting transition.
Finally, limitation on the parameter $\alpha = l_{min}^{2 }/x^{2
}$ is inferred from the relation
$$Sp[\rho x^{2}]-Sp[\rho
a]Sp[\rho x] \simeq l^{2}_{min}\;\; or\;\; Sp[\rho]-Sp^{2}[\rho]
\simeq l^{2}_{min}/x^{2},$$
that follows from (\ref{U6N}) with the
use of the R-procedure \cite{shalyt4},\cite{shalyt5}.
\\According to point 5, $<1>_{\alpha}=Sp[\rho(\alpha)]$. Therefore
for any scalar quantity $f$ we have $<f>_{\alpha}=f
Sp[\rho(\alpha)]$. In particular, the mean value
$<[x_{\mu},p_{\nu}]>_{\alpha}$ is equal to
\\
$$<[x_{\mu},p_{\nu}]>_{\alpha}= i\hbar\delta_{\mu,\nu}
Sp[\rho(\alpha)]$$
\\
We denote the limit $\lim\limits_{\alpha\rightarrow
0}\rho(\alpha)=\rho$ as the density matrix. Evidently, in the
limit $\alpha\rightarrow 0$ we return to QM.
\renewcommand{\theenumi}{\Roman{enumi}}
\renewcommand{\labelenumi}{\theenumi.}
\renewcommand{\labelenumii}{\theenumii.}

It should be noted that:

\begin{enumerate}
\item The above limit covers both Quantum
and Classical Mechanics. Indeed, since $\alpha\sim L_{p}^{2 }/x^{2
}=G \hbar/c^3 x^{2}$, we obtain:
\begin{enumerate}
\item $(\hbar \neq 0,x\rightarrow
\infty)\Rightarrow(\alpha\rightarrow 0)$ for QM;
\item $(\hbar\rightarrow 0,x\rightarrow
\infty)\Rightarrow(\alpha\rightarrow 0)$ for Classical Mechanics;
\end{enumerate}
\item As a matter of fact, the deformation parameter $\alpha$
should assume the value $0<\alpha\leq1$.  As seen from
(\ref{U2}), however, $Sp[\rho(\alpha)]$ is well defined only for
$0<\alpha\leq1/4$. That is if $x=il_{min}$ and $i\geq 2$, then
there is no any problem.  At the point of $x=l_{min}$ there is
a singularity related to the complex values following from
$Sp[\rho(\alpha)]$ , i.e. to the impossibility of obtaining a
diagonalized density pro-matrix at this point over the field of
real numbers. For this reason definition 1 has no sense at the
point $x=l_{min}$.
\item We consider possible solutions for (\ref{U1}).
For instance, one of the solutions of (\ref{U1}), at least to the
first order in $\alpha$, is $$\rho^{*}(\alpha)=\sum_{i}\alpha_{i}
exp(-\alpha)|i><i|,$$ where all $\alpha_{i}>0$ are independent of
$\alpha$ and their sum is equal to 1. In this way
$Sp[\rho^{*}(\alpha)]=exp(-\alpha)$. We can easily verify that
\begin{equation}\label{U3}
Sp[\rho^{*}(\alpha)]-Sp^{2}[\rho^{*}(\alpha)]=\alpha+O(\alpha^{2}).
\end{equation}
 Note that in the momentum representation $\alpha\sim p^{2}/p^{2}_{pl}$,
where $p_{pl}$ is the Planck momentum. When present in the matrix
elements, $exp(-\alpha)$ can damp the contribution of great
momenta in a perturbation theory.
\end{enumerate}

\section{Entropy Density Matrix and Information Loss Problem}
In \cite{shalyt5} the authors were too careful, when introducing
for density pro-matrix $\rho(\alpha)$ the value $S_{\alpha}$
generalizing the ordinary statistical entropy:
\\
 $$S_{\alpha}=-Sp[\rho(\alpha)\ln(\rho(\alpha))]=
 -<\ln(\rho(\alpha))>_{\alpha}.$$
\\
In \cite{shalyt4},\cite{shalyt5} it was noted that $S_{\alpha}$
means of the entropy density   on a unit  minimum area depending
on the scale. In fact a more general concept accepts the form of
the entropy density matrix
\cite{shalyt13},\cite{shalyt13s},\cite{shalyt20}:
\begin{equation}\label{U4}
S^{\alpha_{1}}_{\alpha_{2}}=-Sp[\rho(\alpha_{1})\ln(\rho(\alpha_{2}))]=
-<\ln(\rho(\alpha_{2}))>_{\alpha_{1}},
\end{equation}
where $0< \alpha_{1},\alpha_{2}\leq 1/4.$
\\ $S^{\alpha_{1}}_{\alpha_{2}}$ has a clear physical meaning:
the entropy density is computed  on the scale associated with the
deformation parameter $\alpha_{2}$ by the observer who is at a
scale corresponding to the deformation parameter $\alpha_{1}$.
Note that with this approach the diagonal element
$S_{\alpha}=S_{\alpha}^{\alpha}$,of the described matrix
$S^{\alpha_{1}}_{\alpha_{2}}$ is the density of entropy measured
by the observer  who is at the same scale  as the measured object
associated with the deformation parameter $\alpha$. In
\cite{shalyt5} Section 6 such a construction was used implicitly
in derivation of the semiclassical Bekenstein-Hawking formula for
the Black Hole entropy:

a) For the initial (approximately pure) state
\\
$$S_{in}=S_{0}^{0}=0$$
\\
b) Using the exponential ansatz(\ref{U3}),we obtain:
\\
$$S_{out}=S^{0}_{\frac{1}{4}}=-<ln[exp(-1/4)]\rho_{pure}>=-<\ln(\rho(1/4))>
=\frac{1}{4}.$$
\\
So increase in the entropy density for an external observer
at the large-scale limit is 1/4. Note that increase of
the entropy density  (information loss) for the
observer that is crossing the horizon of the black hole's events
and moving with the information flow to singularity will be smaller:
\begin{eqnarray}
S_{out}=S_{\frac{1}{4}}^{\frac{1}{4}}=-Sp(exp(-1/4)
ln[exp(-1/4)]\rho_{pure}) \nonumber \\
=-<\ln(\rho(1/4))>_{\frac{1}{4}} \approx 0.1947 \nonumber
\end{eqnarray}
It is clear that this fact may be interpreted as follows: for the
observer moving together with information its loss can  occur only
at the transition to smaller scales, i.e. to greater deformation
parameter $\alpha$.
\\Now we consider the general Information Problem.
Note that with the classical Quantum Mechanics (QM) the entropy
density matrix $S^{\alpha_{1}}_{\alpha_{2}}$ (\ref{U4}) is reduced
only to one element $S_{0}^{0}$ and so we can not test anything.
Moreover, in previous works relating the quantum mechanics of
black holes and information paradox \cite{Hawk1},\cite{r18,r19}
the initial and final states when a particle hits the
 hole are treated proceeding from different theories(QM and QMFL respectively):
\\
\\
(Large-scale limit, QM,
 density matrix) $\rightarrow$ (Black Hole, singularity, QMFL,
density pro-matrix),
\\
\\
Of course in this case any conservation of information is
impossible as these theories are based on different concepts of entropy.
Simply saying, it is incorrect to compare the entropy interpretations
of two different theories (QM and QMFL,
where this notion is originally differently understood.
So the chain above must be symmetrized by accompaniment
of the arrow on the left ,so in an ordinary situation we have a
chain:
\\
\\
(Early Universe, origin singularity, QMFL, density pro-matrix)
$\rightarrow$
\\ (Large-scale limit, QM,
 density matrix)$\rightarrow$ (Black Hole, singularity, QMFL,
density pro-matrix),
\\
\\
So it's more correct to compare entropy close to the initial and final
(Black hole) singularities. In other words, it is necessary to
take into account not only the state, where information disappears,
but also that whence it appears. The question arises, whether the
information is lost in this case for every separate observer. For
the event under consideration this question sounds as follows: are the
entropy densities S(in) and S(out) equal for every separate observer?
It will be shown that in all conceivable cases they are equal.

1) For the observer in the large-scale limit (producing
measurements in the semiclassical approximation) $\alpha_{1}=0$
\\
\\
$S(in)=S^{0}_{\frac{1}{4}}$ (Origin singularity)
\\
\\
$S(out)=S^{0}_{\frac{1}{4}}$ (Singularity in Black Hole)
\\
\\
So $S(in)=S(out)=S^{0}_{\frac{1}{4}}$. Consequently, the initial
and final densities of entropy are equal and there is no any
information loss.
\\
2) For the observer moving together with the information flow in
the general situation  we have the chain:
\\
$$S(in)\rightarrow S(large-scale)\rightarrow S(out),$$
\\
where $S(large-scale)=S^{0}_{0}=S$. Here $S$ is the ordinary entropy
at quantum mechanics(QM), but
$S(in)=S(out)=S^{\frac{1}{4}}_{\frac{1}{4}}$,value considered in
QMFL. So in this case the initial and final densities of entropy are
equal without any loss of information.
\\
3) This case is a special case of 2), when we do not come out of
the early Universe considering the processes with the participation
of black mini-holes only. In this case the originally specified
chain becomes shorter by one Section:
\\
\\
(Early Universe, origin singularity, QMFL, density
pro-matrix)$\rightarrow$ (Black Mini-Hole,
singularity, QMFL, density pro-matrix),
\\
\\
and member $S(large-scale)=S^{0}_{0}=S$ disappears at the
corresponding chain of the entropy density associated with
the large-scale consideration:
\\
$$S(in)\rightarrow S(out),$$
\\
It is, however, obvious that in case
$S(in)=S(out)=S^{\frac{1}{4}}_{\frac{1}{4}}$ the density of
entropy is preserved. Actually this event was mentioned in
\cite{shalyt5},where from the basic principles it has been found
that black mini-holes do not radiate, just in agreement with the
results of other authors \cite{r14}-\cite{r17}.
\\ As a result, it's possible to write briefly
\\
$$S(in)=S(out)=S^{\alpha}_{\frac{1}{4}},$$
\\
where $\alpha$ - any value in the interval $0<\alpha\leq 1/4.$
\\ It should be noted that in terms of deformation the Liouville's
equation (Section 4 \cite{shalyt5}) takes the form:
\\
$$\frac{d\rho}{dt}=\sum_{i}
\frac{d\omega_{i}[\alpha(t)]}{dt}|i(t)><i(t)|-i[H,\rho(\alpha)]=
\\d[ln\omega(\alpha)]\rho (\alpha)-i[H,\rho(\alpha)].$$
\\
The main result of this Section is a necessity to
account for the member $d[ln\omega(\alpha)]\rho (\alpha)$,deforming
the right-side expression of $\alpha\approx 1/4$.

\section {Entropy Bounds, Entropy Density and \\Holographic Principle}
In the last few years Quantum Mechanics of black holes has been
studied under the assumption that GUR are valid
\cite{r14},\cite{r15},\cite{r17}. As a result of this approach, it
is indicated that the evaporation process of a black hole gives a
stable remnant with a mass on the order of the Planck's $M_{p}$.
However, J.Bekenstein in \cite{bek1} has credited such an approach
as problematic, since then the objects with dimensions on the
order of the Planck length $\sim 10^{-33}cm$ should have very
great entropy thus making problems in regard to the entropy bounds
of the black hole remnants \cite{bek2}.
\\In connection with this remark of J.Bekenstein \cite{bek1}
the following points should be emphasized:
\\ I. An approach proposed in \cite{shalyt11},\cite{shalyt13}
and in the present paper gives a deeper insight into the cause of
high entropy for Planck's black hole remnants, namely: high
entropy density that by this approach at Planck scales takes place
for every fixed observer including that on a customary scale, i.e.
on $\alpha\approx 0$. In \cite{shalyt13} using the exponential
ansatz (Section 3) it has been demonstrated how this density can
increase in the vicinity of the singularities with
\\
$$S_{in}=S_{0}^{0}\approx 0$$
\\
up to\\ $$S_{out}=S^{0}_{\frac{1}{4}}=-<ln[exp(-1/4)]\rho_{pure}>
=-<\ln(\rho^{*}(1/4))> =\frac{1}{4}.$$
\\
when the initial state measured by the observer is pure.
\\As demonstrated in \cite{shalyt11},\cite{shalyt13}, increase
in the entropy density  will be realized also for the observer
moving together with the information flow:
$S_{out}=S^{\frac{1}{4}}_{\frac{1}{4}}>S_{0}^{0}$, though to a
lesser extent than in the first case. Obviously, provided the
existing solutions for (\ref{U1}) are different from the
exponential ansatz, the entropy density for them
$S^{0}_{\alpha_{2}}$ will be increasing as compared to $S_{0}^{0}$
with a tendency of $\alpha_{2}$ to 1/4.
\\II. In works of J.Bekenstein, \cite{bek2} in particular,
a "universal entropy bound" has been used \cite{bek3}:
\begin{equation}\label{UBec}
S\leq 2\pi MR/\hbar,
\end{equation}
where $M$ is the total gravitational mass of the matter and $R$ is
the radius of the smallest sphere that barely fits around a
system. This bound is, however, valid for a weakly gravitating
matter system only. In case of black hole remnants under study it
is impossible to assume that on Planck scales we are concerned
with a weakly gravitating matter system, as in this case the
transition to the Planck's energies is realized where
quantum-gravitational effects are appreciable, and within the
proposed paradigm  parameter $\alpha\approx 0$ is changed by the
parameter $\alpha>0$ or equally QM is changed by QMFL.
\\
\\III.This necessitates mentioning of the recent findings of R.Bousso
\cite{bou1},\cite{bou2}, who has derived the Bekenstein's
"universal entropy bound" for a weakly gravitating matter system,
and among other things in flat space, from the covariant entropy
bound \cite{bou3} associated with the holographic principle of
Hooft-Susskind \cite{hol1},\cite{hol2},\cite{hol3}.
\\ Also it should be noted that the approach proposed in
\cite{shalyt13},\cite{shalyt5} and in the present paper is
consistent with the holographic principle \cite{hol1}-\cite{hol3}.
Specifically, with the use of this approach one is enabled to
obtain the entropy bounds for nonblack hole objects of L.Susskind
\cite{hol2}. Of course, in (\cite{shalyt5}, Section 6) and
(\cite{shalyt13}, Section 4) it has been demonstrated, how a
well-known semiclassical Bekenstein-Hawking formula for black hole
entropy may be obtained using the proposed paradigm. Then we can
resort to reasoning from \cite{hol2}: "using gedanken experiment,
take a neutral non-rotating spherical object containing entropy
$S$ which fits entirely inside a spherical surface of the area
$A$, and it to collapse to black hole". Whence
\begin{equation}\label{USuss}
S\leq \frac{A}{4l^{2}_{p}}.
\end{equation}
Note also that the entropy density matrix
$S^{\alpha_{1}}_{\alpha_{2}}$ by its definition
\cite{shalyt11},\cite{shalyt13} falls into 2D objects, being
associated with $l^{2}_{min}\sim l^{2}_{p}$ \cite{shalyt5} and
hence implicitly pointing to the holographic principle.
\\Qualitative analysis
performed in this work reveals that the Information Loss Problem
in black holes with the canonical problem statement
\cite{Hawk1},\cite{r18},\\ \cite{r19} suggests in principle
positive solution within the scope of the proposed method -
high-energy density matrix deformation. Actually, this problem
necessitates further (now quantitative) analysis. Besides, it is
interesting to find direct relations between the described methods
and the holographic principle. Of particular importance seems a
conjecture following from \cite{bou2}:
\\is it possible to derive GUR for high energies
(of strong gravitational field) with the use of the covariant
entropy bound \cite{bou3} in much the same manner as R.Bousso
\cite{bou2} has developed the Heisenberg uncertainty principle for
the flat space?

\section{Quantum corrections to black hole entropy. Heuristic
approach}
This paper presents certain results pertinent to the
application of the above methods in a Quantum Theory of Black
Holes. Further investigations are still required in this respect,
specifically for the complete derivation of a semiclassical
Bekenstein-Hawking formula for the Black Hole entropy, since in
Section 2 the treatment has been based on the demonstrated result:
a respective number of the degrees of freedom is equal to $A$,
where $A$ is the surface area of a black hole measured in Planck's
units of area $L_{p}^{2}$ (e.g.\cite{Kh1},\cite{Kh2}). Also it is
essential to derive this result from the basic principles given in
this paper.
\\ As indicated in papers \cite{shalyt14},\cite{shalyt20},
the calculation procedure in the described theory may be reduced
to a series expansion in terms of $\alpha$ parameter and to
finding of the factors for the ever growing powers of this
parameter, that may be considered in some cases as the calculation
of quantum corrections. Specifically, this approach to calculation
of the quantum correction factors may be used in the formalism for
density pro-matrix (Definition 1 of Section 1). In this case, the
value $S^{0}_{\alpha}$  point b), in Section 2 may be written in
the form of a series
\begin{equation}\label{U28L}
S^{0}_{\alpha}=\alpha+a_{0}\alpha^{2} +a_{1}\alpha^{3}+... .
\end{equation}
As a result, a measurement procedure using the exponential ansatz
may be understood as the calculation of factors
$a_{0}$,$a_{1}$,... or the definition of additional members in the
exponent "destroying" $a_{0}$,$a_{1}$,... \cite{shalyt13}. It is
easy to check that close to the singularity $\alpha=1/4$ the
exponential ansatz gives $a_{0}=-3/2$, being coincident with the
logarithmic correction factor for the black hole entropy
\cite{r22}.
\\However, by the proposed approach - density matrix
deformation at Planck's scales - the quantum entropy receives a
wider and more productive interpretation due to the notion of
entropy density matrix introduced in
\cite{shalyt13},\cite{shalyt13s},\cite{shalyt20},\cite{shalyt11}
and Section 3. Indeed, the value
$S^{\alpha_{1}}_{\alpha_{2}}=-Sp[\rho(\alpha_{1})\ln(\rho(\alpha_{2}))]$
may be considered as a series of two variables $\alpha_{1}$ and
$\alpha_{2}$. Fixing one of them, e.g. $\alpha_{1}$, it is
possible to expand the series in terms of $\alpha_{2}$ parameter
and to obtain the quantum corrections to the main result as more
and more higher-order terms of this series. In the process,
(\ref{U28L}) is a partial case of the approach to $\alpha_{1}=0$
and $\alpha_{2}$ close to 1/4.
\section{Conclusion}
Thus, in this paper it is demonstrated that the developed approach
to study a quantum theory of the Early Universe - density matrix
deformation at Planck's scales - leads to a new method of studying
the black hole entropy its quantum aspects including. Despite the
fact that quite a number of problems require further
investigation, the proposed approach seems a worthy contribution:
first, it allows for the direct calculations of entropy and,
second, the method enables its determination proceeding from the
basic principles with the use of the density matrix. Actually, we
deal with a substantial modification of the conventional density
matrix known from the quantum mechanics - deformation at Planck's
scales or minimum length scales. Moreover, within this approach it
is possible to arrive at a very simple derivation and physical
interpretation for the Bekenstein-Hawking formula of black hole
entropy in a semi-classical approximation.  Note that the proposed
approach enables one to study the information problem of the
Universe proceeding from the basic principles and two types of the
existing quantum mechanics only: QM that describes nature at the
well known scales and QMFL at Planck's scales.  The author is of
the opinion that further development of this approach will allow
to research the information problem in greater detail. Besides, it
is related to other methods, specifically to the holographic
principle, as the entropy density matrix studied in this work is
related to the two-dimensional object. Also, it should be noted
that there is an interesting new approach to calculation of the
black hole entropy on the basis of GUR as well \cite{Med1}.
\\ To conclude, it should be noted that an important problem
of the extremal black holes was beyond the scope of this paper. In
the last decade this problem has, however, attracted much
attention in connection with a string theory and quantum
gravitation(e.g., \cite{Mal1}--\cite{Hor}). Specifically, the
Bekenstein-Hawking formula has been proved for this case by
different methods. The author is hopeful that the approaches
proposed in this paper may be developed further to include this
very important problem as well.


\end{document}